\documentclass[peer-review]{fa2026}

\title{The role of sound and auditory displays in telescope control rooms: a pilot study}

\author[1]{Anita Zanella}
\author[2, 3]{Sara Lenzi}
\author[4]{Sandra Pauletto}
\author[5]{Wolfgang Aigner}
\correspondingauthor{anita.zanella@inaf.it}{A. Zanella et al.}

\affil[1]{Istituto Nazionale di Astrofisica - Osservatorio di astrofisica e scienza dello spazio, Italy}
\affil[2]{Faculty of Engineering, University of Deusto, Spain}
\affil[3]{Ikerbasque Basque Foundation for Science, Spain}
\affil[4]{Department of Media Technology and Interaction Design, KTH Royal Institute of Technology, Sweden}
\affil[5]{Institute of Creative\textbackslash Media/Technologies, University of Applied Sciences St. Pölten, Austria}

\begin{document}
\maketitle

\begin{abstract}
Astronomers are often stereotyped as gazing at the stars. Today, they mainly visually inspect digital data they receive from telescopes in remote observatories. One of the most complex is the Very Large Telescope (VLT) in Chile. Multiple streams of information converge in the VLT control room, where telescopes and instruments are managed, with operators handling numerous tasks on dozens of screens in a challenging environment, marked by harsh geography, intense work rhythms, and isolation. Sound and sonification can represent an efficient novel means for data monitoring with efficiency and operator wellness in mind. We present insights from a qualitative pilot study aimed at assessing the current use of sound in the VLT control room to identify directions for the design of new auditory displays. Based on the analysis of questionnaires and interviews with VLT personnel, we identify the key characteristics of the current use of sound and describe opportunities for improved design strategies to integrate auditory displays in telescope control rooms to support decision-making, improve situational awareness, and reduce cognitive load. 

\end{abstract}


\section{Introduction}\label{sec:introduction}

Astronomers are generally portrayed as individuals gazing at the stars. While this is no longer the case for professional astronomers, data visualisation, namely the visual display of information through images and graphs, remains the primary method for exploring and presenting astronomical data. Nowadays, astronomers in research centres around the world primarily receive data from observatories situated in remote locations, far from light pollution. These observatories are staffed by diverse teams, including astronomers, mechanical and software engineers, and telescope and instrument operators (TIOs), who handle tasks such as data collection, quality assurance, maintenance, and real-time monitoring of weather conditions. Information on these activities is displayed as data visualizations on screens. However, this reliance on visual information exposes operators to large volumes of data, with the risk of cognitive overload, a phenomenon linked to increased burnout and other health problems \cite{Girard2008, Mahapatra2018}, decreased work performance \cite{Hunter2008, Varisco2021}, and slower response times in critical situations \cite{Johnson2017}. 

In this work, we investigate how auditory information is currently used to complement visual data and how operators perceive such sounds, with the aim of informing future audio-visual designs that may enhance work performance, support decision-making, improve situational awareness, and reduce cognitive load.
We use the control room of the Very Large Telescope (VLT), one of the world's most advanced and complex ground-based astronomical observatory, as our ``living lab.'' The VLT is located in the Atacama Desert (Chile) and is managed by the European Southern Observatory (\href{https://www.eso.org/public/about-eso/}{ESO}). Observation time at the VLT is extremely precious, and its effective use is crucial: each night of observations cost about 150,000 EUR. Multiple streams of information converge in this hub, with dozens of astronomers and TIOs managing various tasks from desks saturated by more than ten screens, within an open office space where other activities (chatting, phone calls, people moving around) also happen contextually. 
The situation is bound to become even more challenging. ESO is currently building the Extremely Large Telescope (ELT) that will start operations in 2030 and, with its 39-metre mirror, will be the largest telescope ever built. The ELT will be operated from the current VLT control room, and, following the current standard approach, data will be displayed on even more screens. The ELT will produce tens of terabytes of multi-dimensional, and possibly non-stationary (i.e., requiring real-time or pseudo-real-time monitoring) data each night. 

Current communication at the VLT relies primarily on visual means, with auditory alarms alerting operators to emergencies. These alarms have not been intentionally designed; instead, they have been chosen over the years based on the (sometimes bizarre!) preferences of individual operators\footnote{Examples can be heard at this link: \href{https://tinyurl.com/ESOAlarmSounds}{https://tinyurl.com/ESOAlarmSounds}}. Sounds include iconic sentences extracted from famous sci-fi movies, as well as sound effects (e.g., WC flushing) and more typical alarms, such as sirens. In terms of function, these can be considered as auditory cues, which provide real-time feedback to specific actions, and traditional alarms, which inform of an emergency. Such cues and alarms are played after the event they represent has occurred. Thus, they limit the agency of the operator who, in case of emergency, can only react to mitigate damages.

\section{Background}
\label{sec:background}

Data sonification and auditory displays, namely the use of non-speech audio to represent data and their behaviour over time, is a growing research field that moves beyond visual representations \cite{Lindborg2024}. Sound offers several advantages for data communication. Changes in acoustic patterns are easily detected by the human ear \cite{Neuhoff2011}. Sound composition is inherently multivariate, and the human ear can distinguish remarkably well multiple sonic characteristics (e.g. pitch, loudness, timbre) simultaneously \cite{Chion2015}. Lastly,  auditory signals are superior for temporal discrimination, the perception of rapidly changing signals, and the detection of background anomalies while attention is directed elsewhere \cite{diaconescu2013visual}. However, sonification has struggled to gain traction outside academia and has not yet become a mainstream method of human-data communication \cite{Supper2012}. The reasons are manifold. Unlike data visualisation, which benefits from established frameworks and epistemic instruments \cite{Cairo2012, Munzner2025}, the auditory display community has yet to reach consensus on design strategies and data-to-sound mappings \cite{Brazil2009, Dubus2013}. Research on these topics is limited, and sonification solutions often neglect the aesthetic quality needed for real-world use \cite{Vickers2011}. Lastly, a critical area is the lack of training protocols for users unfamiliar with sonification \cite{Lenzi2024}, which, on the contrary, appears to be extremely promising in reducing information overload \cite{Stadin2020}. Currently, both education and professional training methods focus on visualisation, and we are not trained to listen attentively for the purpose of gaining and analysing complex information, a skill required to effectively use sonification \cite{Scaletti2017}. So far, researchers have addressed these issues on a case-by-case basis, often reinventing the wheel and on a small scale (e.g., engaging few potential users in laboratory trials over a limited period), with a generic sense of under-achievement, if not blunt failure \cite{Neuhoff2019}. In this context, user-centred and participatory methods \cite{abras2004user,schuler1993participatory}, which aim to include diverse stakeholders throughout the design process, are key to the development of successful solutions. Within the field of astronomy, participatory and interdisciplinary workshops have been successfully used to develop innovative sonification solutions \cite{misdariis2023audible}. A follow-up study conducted two years later confirmed the impact of such participatory activities, as the design concepts generated during these workshops were effectively implemented in audiovisual tools for astronomy and education \cite{pauletto2025sound}. 
This study represents an initial effort to involve astronomers and TIOs in the design of novel auditory display solutions for the telescope control room.  Through questionnaires and interviews, we aimed at define current sound usage in the control room while taking into account anticipated developments in telescope control room environments.

\begin{figure*}[t!]
    \centering
    \includegraphics[width=0.7\linewidth]{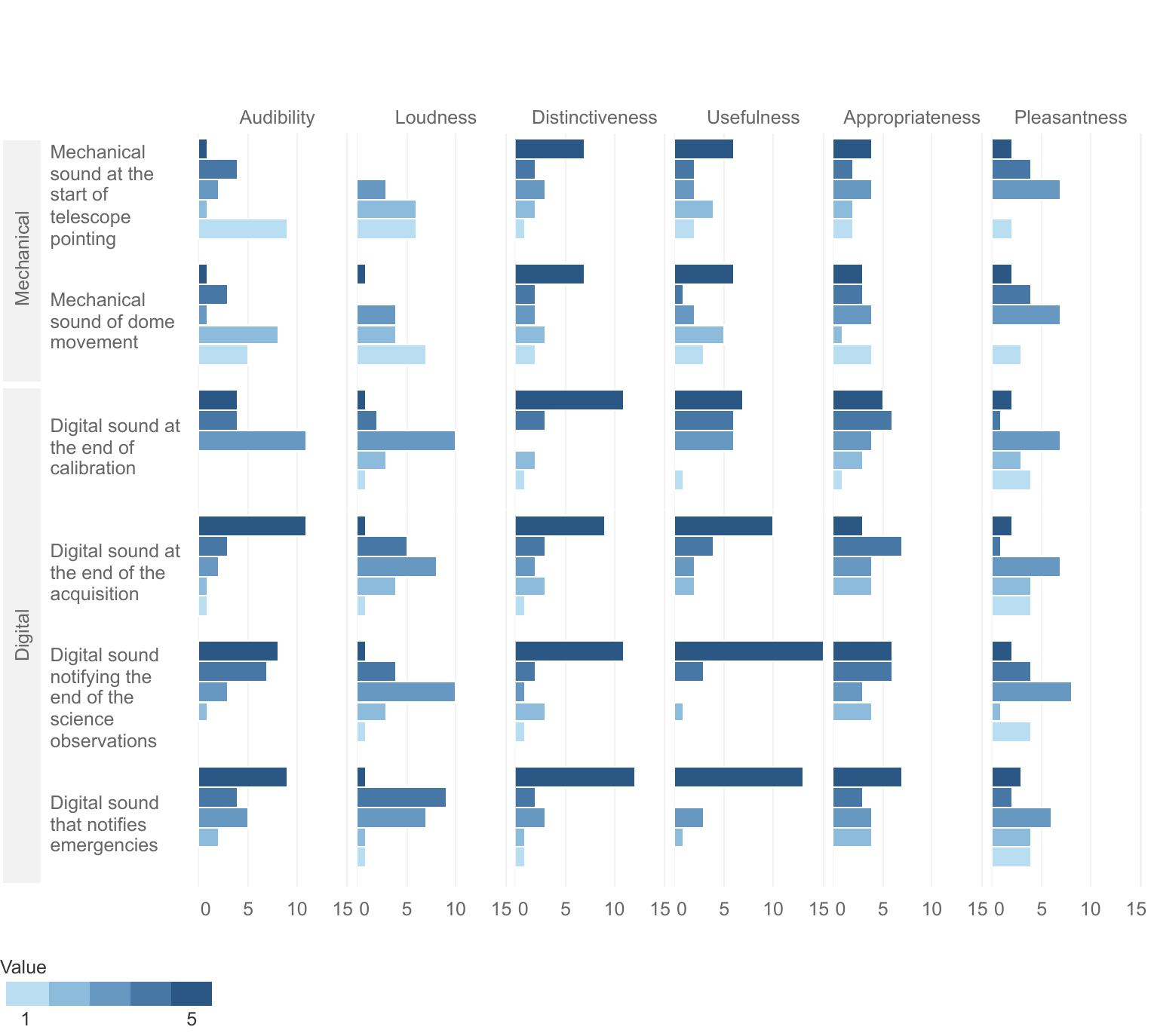}
        \includegraphics[width=0.28\linewidth]{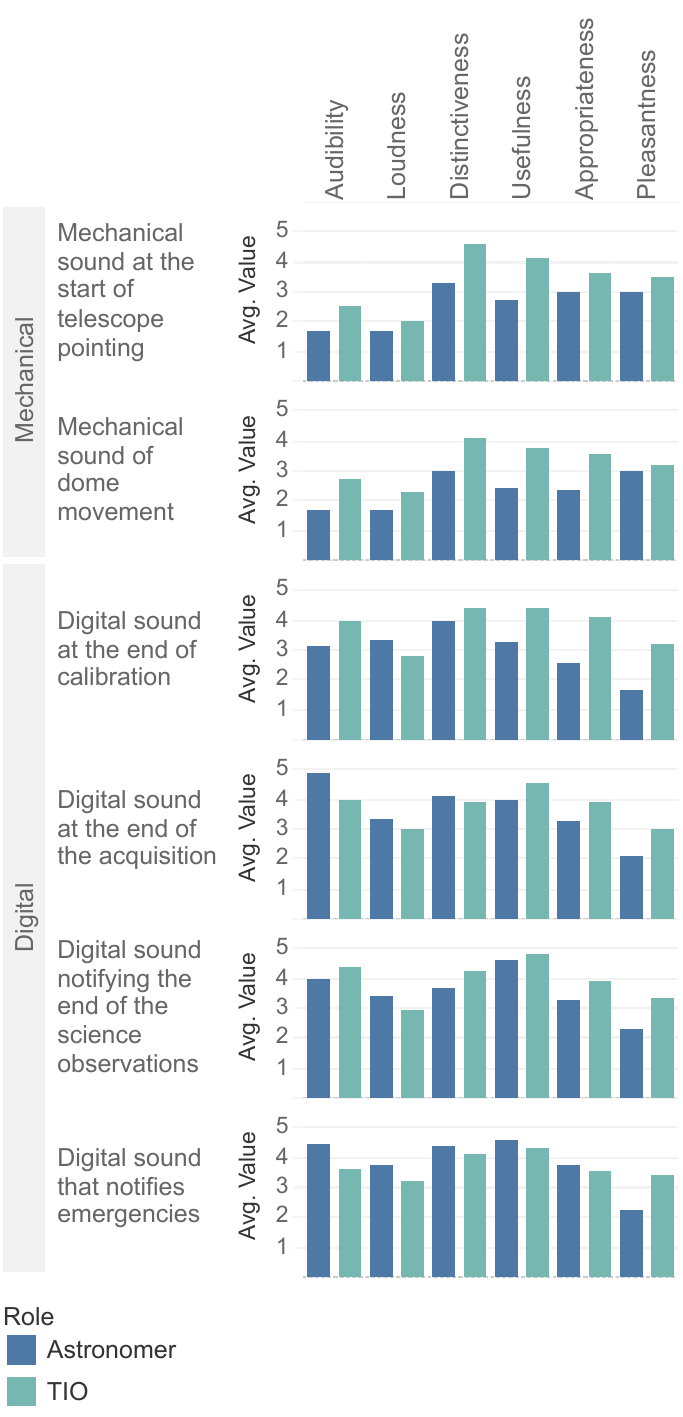} 
        \caption{Bar charts showing the perceived characteristics of different sounds present in the control room. \textbf{Left panel}: darker colors indicate higher degrees of agreement on a scale from 1 (``not at all'', teal) to 5 (``very much'', blue) to the different sound qualities shown in columns. \textbf{Right panel}: comparison of the average scores assigned to the different sounds by astronomers (blue) and TIOs (teal).}
    \label{fig:barchart_astroTIOs}
\end{figure*}

\section{Methods}
\label{sec:methods}

We carried out our research using both a written questionnaire and dedicated interviews. In the following, we summarize insights from both approaches. In particular, we distinguish among mechanical and digital sounds: the first are sounds directly produced by the instrumentation (e.g., the rustle of the telescope moving); the latter are artificial sounds emitted by speakers on the monitor. We also distinguish participants according to their professional role: TIOs generally perform tasks related to the telescope (e.g., opening and closing the dome, pointing the telescope, etc.), while astronomers decide what observations to perform, operate the instruments, assess the quality of the data, and are in charge of writing the night log.

\subsection{Questionnaire}
\label{subsec:questionnaire}
Participants were recruited via their work email addresses and responded on a voluntary basis. We received 20 responses (6 women, 14 men, aged 31-54): 9 from astronomers, 10 from TIOs, and 1 unknown.
The questionnaire was in English, created with Google Forms, and structured into three sections\footnote{The survey questions can be read at this link: \href{https://docs.google.com/forms/d/e/1FAIpQLSfpKhulMnePATBSvaOX32kiWRsKnVDkY0ApWcRNfRhkFO7Qtw/viewform?usp=sharing&ouid=107596044837984465667}{\color{blue}link}}: 

\textit{Section 1 (Process and importance)}: we collected demographic information. We investigated the various tasks that astronomers and TIOs perform over night and the cognitive load that they require. We also included questions about the characteristics of existing sounds.

\textit{Section 2 (Presence and utility)}: we focused on the presence or absence of specific sounds and their perceived utility. This section aimed to map the current auditory landscape of the control room.

\textit{Section 3 (Personal preferences)}: we examined individual habits (e.g., use of headphones, listening to music, and volume preferences) to assess how they interact with work-related auditory stimuli.

\subsection{Interviews}
\label{subsec:interviews}

Three individuals (two women and one man, 35-50 years old) who have worked or are currently working at the VLT were recruited for semi-structured interviews. Two participants are astronomers, while the third is a member of the management team with a background in astronomy. 
The interviews were conducted online, recorded and transcribed using Atlas.ti (version 25.02). They were organized around three main themes: the interviewees' professional and personal relationship with the VLT context; their relationship with the soundscape; and their interaction with data in daily work practices (e.g., acquisition, processing, and visualization).
The primary aim was to understand the broader experience of living and working in the VLT's extreme and isolated environment, with particular attention to how individuals relate to sound both professionally and personally. Additionally, the study explored how data and information are managed at the VLT and how these processes intersect with the soundscape.
This paper focuses on the second theme, namely the relationship with the VLT soundscape and, more specifically, with perceived sound events.

\section{Results}
\label{sec:results}

We adopted both quantitative and qualitative approaches. The quantitative analysis (Section \ref{subsec:quantitative_res}) focused on questionnaire responses measuring participants' perceptions using numerical rating scales (e.g., from 1 ``not at all'' to 5 ``very much''). The qualitative analysis (Section \ref{subsec:qualitative_res}) was based on responses to the semi-structured interview.

\subsection{Quantitative results}
\label{subsec:quantitative_res}

We focused on the main mechanical and digital sounds present in the control room, namely the mechanical sounds produced during telescope pointing and dome movement, and the digital sounds played at the end of instrument calibrations, data acquisition, science observations, and during emergencies. Respondents reported that natural sounds are generally not perceived in the control room, except for the howling of the wind. We investigated six characteristics of these sounds: audibility, loudness, distinctiveness, usefulness, appropriateness, and pleasantness.

Figure \ref{fig:barchart_astroTIOs} shows the results of this analysis. The most clearly audible sounds are those marking the end of data acquisition (78\% rated 4 or 5), emergencies (65\% rated 4 or 5), and the end of science observations (79\% rated 4 or 5). The least audible sounds are those associated with the start of telescope pointing (59\% rated 1 or 2) and dome movement (72\% rated 1 or 2). In general, mechanical sounds appear to be less audible than digital ones.
The sounds are not perceived as particularly loud or quiet. The loudest sound is used for emergencies (average rating 3.5). The quietest sounds are those associated with the start of telescope pointing (average rating 2.3) and dome movement (average rating 2.5). These are also the sounds identified as the least audible.
In general, the sounds are clearly distinguishable, with average ratings ranging from 3.6 to 4.2. The sounds with the lowest averages (dome movement: 3.6; start of telescope pointing: 3.8) are also those with the lowest audibility. Notably, these are both mechanical (vs digital) sounds.
Sounds are deemed useful, in particular those marking the end of data acquisition (67\% rated 4 or 5), emergencies (61\% rated 4 or 5), the end of calibration (68\% rated 4 or 5), and science observations (84\% rated 4 or 5). Similarly, the sounds perceived as most appropriate are those indicating the end of data acquisition (50\% rated 4 or 5), emergencies (50\% rated 4 or 5), the end of calibration (53\% rated 4 or 5), and the end of science observations (58\% rated 4 or 5).
None of the sounds are perceived as particularly pleasant. The average ratings range from 2.8 to 3.2.

\begin{figure}[t!]
    \centering
    \includegraphics[width=\linewidth]{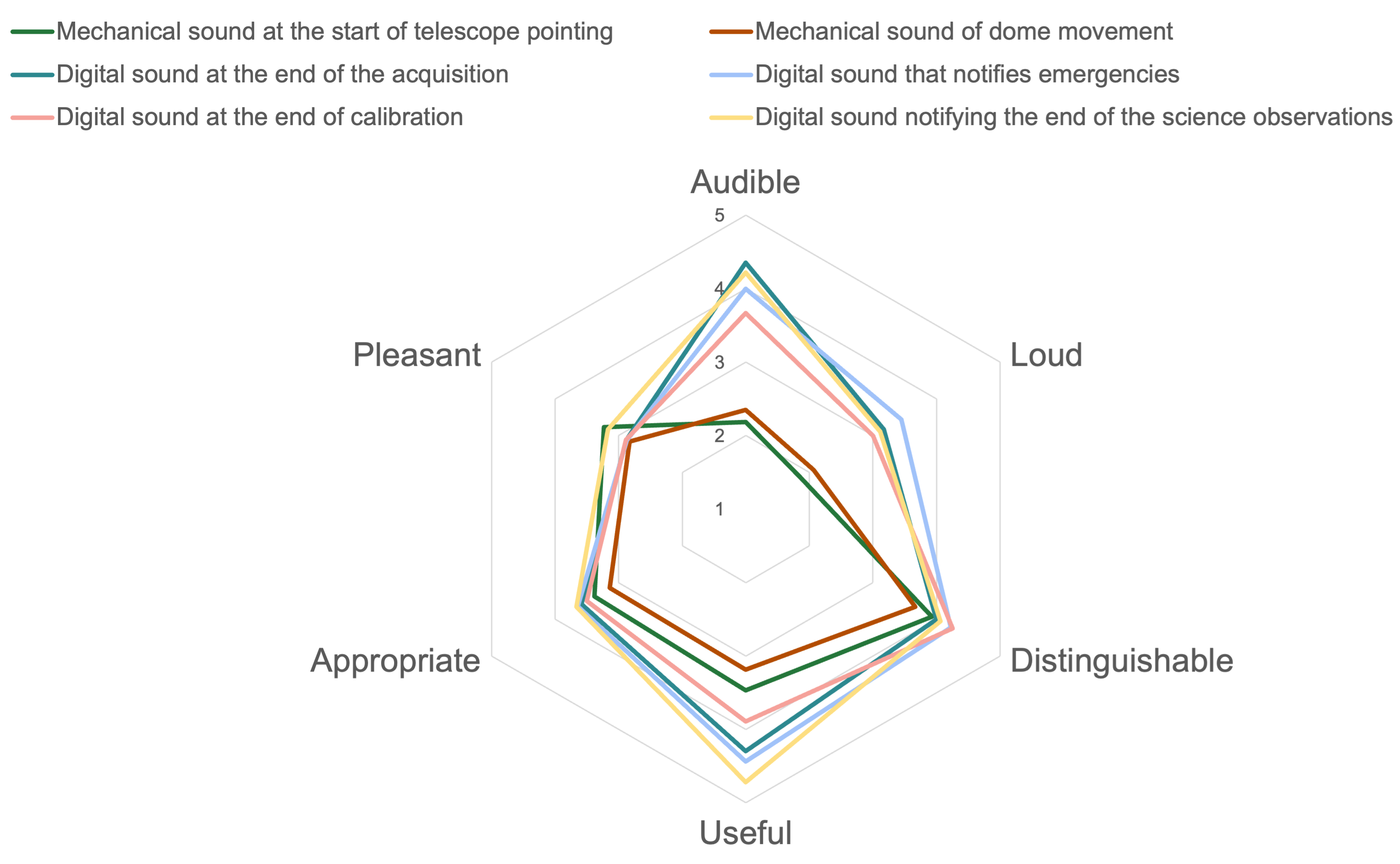}
    \caption{Radar chart showing the average perception of the different sounds (each sound is represented with a different color, as reported in the legend).}
    \label{fig:radar}
    \vspace{-0.5cm}
\end{figure}

As shown in Figure \ref{fig:radar}, there is a clear difference between the ratings of mechanical and digital sounds. Mechanical sounds score lower across all dimensions except pleasantness, which is similar for all. In particular, audibility and loudness receive the lowest scores. Mechanical sounds are also perceived as less useful, less appropriate, and less distinguishable.

Finally, we investigated if sounds are perceived differently by astronomers and TIOs (Figure \ref{fig:barchart_astroTIOs}). In general, TIOs assign higher ratings to mechanical sounds than astronomers do, while ratings for digital sounds  are more similar. Additionally, TIOs perceive the sounds as slightly more pleasant overall (average ratings ranging from 2.9 to 3.7 for TIOs and from 2.7 to 3.1 for astronomers), more appropriate (average ratings ranging from 3.7 to 4.0 for TIOs and from 2.5 to 3.5 for astronomers), and more useful (average ratings ranging from 4.0 to 4.9 for TIOs and from 2.4 to 4.6 for astronomers) than astronomers do.

\subsection{Qualitative results}
\label{subsec:qualitative_res}

\begin{figure}
    \centering
    \includegraphics[width=\linewidth]{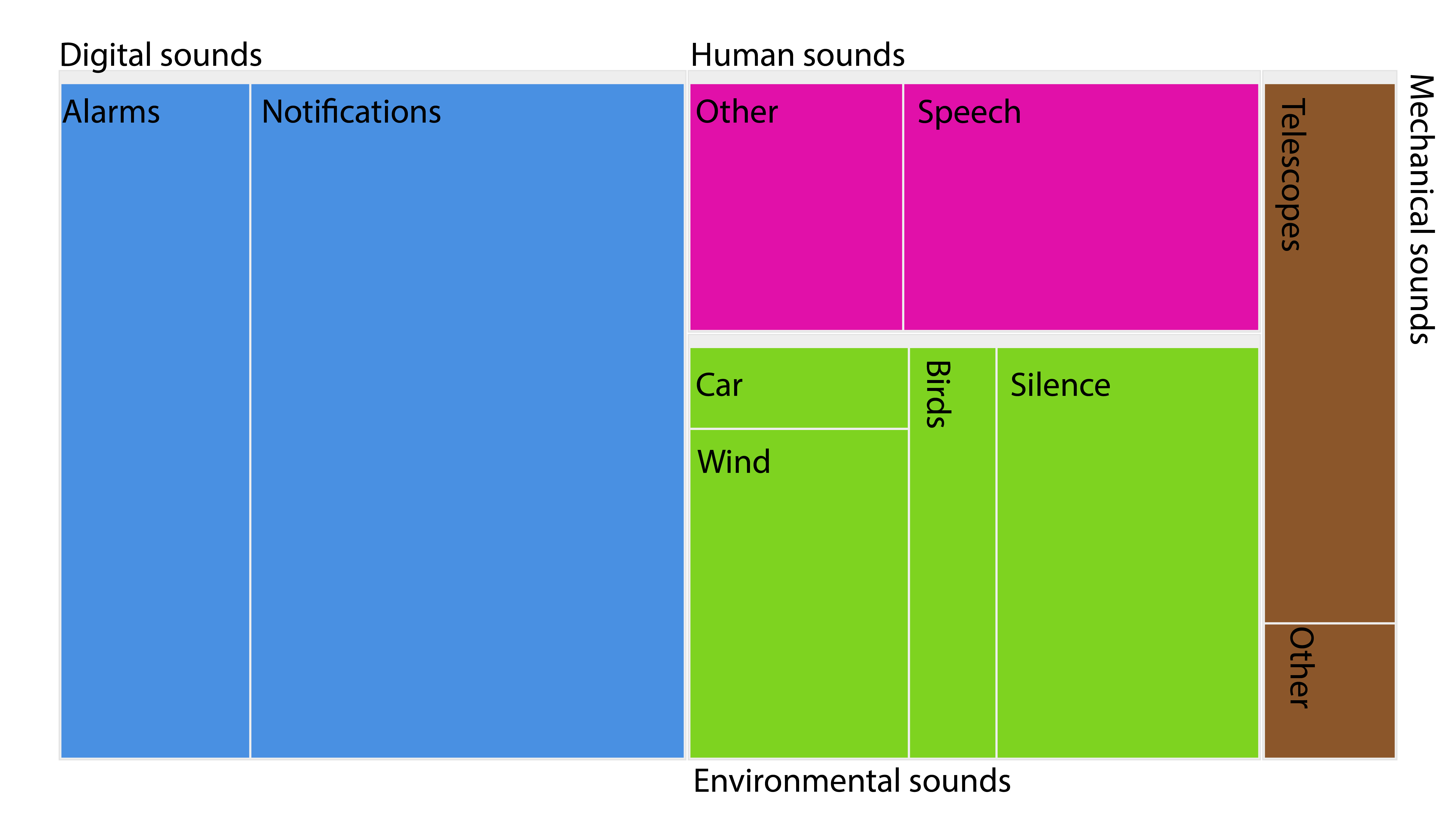}
    \caption{Treemap showing the contributors to the sound categories (\textit{digital} in blue, \textit{human} in pink, \textit{environmental} in green, \textit{mechanical} in brown) and their occurrence in the text.}
    \label{fig:placeholder}
\end{figure}

The interviews were analyzed following the principles of thematic analysis \cite{Braun2012}. Through iterative coding, five themes were identified: The Mountain, The Control Room, Future of the VLT, Human–data Relationships, and The Soundscape.
This study focuses on the soundscape, examining its sub-themes and their relationship to the quantitative results. Figure \ref{fig:placeholder} presents the sound events identified through the thematic analysis and their contribution to the overarching theme of the soundscape, as well as to sub-themes derived by clustering the sound events into higher-level categories: \textit{human} (speech and other human-generated noise), \textit{mechanical }from instruments (telescopes and related equipment), \textit{environmental} (cars, wind, animals), and \textit{digital }(alarms and notifications) sounds.

Human sounds primarily refer to speech within the control room and are only marginally mentioned in relation to other spaces, such as the cafeteria or staff residences. Notably, in shared living areas, human-generated sounds are often perceived as a source of annoyance ("\textit{Sometimes they have parties at the residency, it is an unpleasant sound because I want to sleep}''). In contrast, within the control room colleagues' voices are described as neutral or even positive elements ("\textit{There's people talking in the control room, but it is never too much}").

Environmental sounds outside the control room reflect the desert around the VLT, where silence and wind are the primary soundmarks ("\textit{There is no sound. It's really amazing}'', "\textit{If there's wind, then you have sound. But apart from this, there's no sound}'').

Mechanical sounds are produced by instrumentation, both within and outside the control room. In the interviews, these sounds are described as having both a functional role, such as signaling potential malfunctions ("\textit{The telescopes, every time they move, make a sound. And when they don't make a sound, it's a problem}''), and a personal, affective dimension ("\textit{One of the sounds that always makes me think of Paranal is definitely the cryogenic pumps. It's the exact same sound you hear when you get an MRI scan. So, when I had an MRI and heard that sound, it was almost a source of comfort for me. That hum of the cryogenic pump is very familiar and reassuring to me}'').

A similar dual nature, functional and affective, is evident in the category of digital sounds, which primarily consists of alarms and audio cues in the control room. They "\textit{let us know whether something is not going well or if a task is completed; they are fundamental to our daily work}". For example, "\textit{when an observation ends, a specific sound indicates that it is wrapping up, allowing us to notice it on the computer and proceed with the next steps}". Sound also plays a role in mitigating screen fatigue: "\textit{constantly staring at screens is exhausting, so sound helps track what is happening in real time}".
Some sounds are described as extremely annoying: "\textit{when your telescope is propagating (i.e., projecting laser beams into another telescope's field), the system emits a very loud siren-like sound, extremely annoying, but necessary to alert us quickly; it can be quite startling}". At the same time, others are perceived as amusing. Many cues include speech ("\textit{after each template operation you hear `Aliens approaching'. It's quite funny but everyone has it as a constant background}") or appear to be drawn from everyday sounds ("\textit{the sound at the end of that operation is like a toilet flushing}").

As the VLT control room is a shared space, cues and alarms from one instrument are heard by all operators alike ("\textit{I hear sounds from other telescopes that I should not hear, but we are in the same room...}''). However, while "\textit{most of the sounds are not designed to create a cohesive environment,}'' they still "\textit{serve their purpose effectively. In the end, you learn to ignore those that are not relevant to your task}".

Lastly, auditory cues can help reduce cognitive load during long shifts, since "\textit{you cannot spend the whole night staring at a screen}''. In fact, sound could be leveraged even further: "\textit{It would actually be beneficial to have auditory alerts for such cases, especially for weather data, since weather is very important to us. Currently, we only have visual dashboards for weather}''.

\section{Discussion}
\label{sec:discussion}
TIOs assigned higher average ratings to mechanical sounds than astronomers. One possible explanation is spatial: TIOs are typically positioned closer to the monitors emitting these sounds, rendering them more audible and distinguishable. Additionally, given their operational responsibilities for the dome and telescope, TIOs may attend more closely to such auditory cues, therefore perceiving them as more useful and contextually appropriate. 
This insight is complemented by the interviews where participants (notably, all astronomers) report that mechanical sounds could be extremely insightful although currently "\textit{we don't pay much attention to it, we never really consider that sound, we are not inside the telescope so cannot really hear them}''. These findings suggest that auditory access and relevance may vary across user groups depending on their roles and tasks, and their position within the control room. Consequently, the design of soundscapes could extend beyond the intrinsic characteristics of sounds to also consider factors such as reproduction quality, spatialization, and directionality.

Notably, none of the sound categories were evaluated as neither pleasant or unpleasant. This outcome may reflect users' habituation to the auditory environment and/or a perceived lack of agency in influencing sound design within the control room. However, interviewees repeatedly defined some digital sounds as \textit{funny} as well as some mechanical sounds as representing a long-lasting affective bond with the VLT (see Section \ref{subsec:qualitative_res}). This insight stems from the user-centered approach employed in this study and could be explicitly leveraged to create future functional sounds that can also increase engagement, thus potentially improving both the performance and well-being of operators.

A distinct social aspect of the VLT sounds seems to emerge from this study: several of the digital sounds in the control room are from iconic films (e.g. 2001 Space Odyssey), both playing with explicit cultural references and helping break the seriousness of the work, as reported in the interviews. Notably, the use of speech contrasts with the typical definition of data sonification \cite{Hermann2011}. This choice may be motivated less by considerations of appropriateness and more by a need to mitigate feelings of isolation and to strengthen social bonds in a challenging and silent environment.

\section{Conclusion}
\label{sec:conclusions}
Through the engagement of domain experts (astronomers and TIOs), this study investigated the role of sound within the control room of the VLT, conceptualized as a complex socio-technological environment. The findings indicate that mechanical and digital sounds are perceived in different ways, with mechanical sounds receiving consistently lower scores across most evaluative dimensions. Furthermore, TIOs and astronomers exhibit divergent perceptual responses, highlighting the importance of accounting for user-specific needs in the design of future auditory systems.

In summary, the sounds of telescope control rooms have both functional and social aspects and they should be designed in collaboration with users, having operators and tasks in mind; mechanical sounds, if intentionally augmented, can be more meaningful; and personalized solutions (e.g. through spatialization) could reduce sonic clutter. Furthermore, future research would benefit from acoustic measurements within the observatory to systematically assess the effects of sonic signals on perception, as well as, where feasible, longitudinal studies involving larger participant groups.

\small
\paragraph{Acknowledgments} AZ thanks A. Minozzi for data collection, L. Guiotto Nai Fovino and M. Grassi for useful discussions, A. Mehner for sharing the survey with ESO's employees, and all astronomers and TIOs who responded. SL work is partially funded by the EU MSCA Cofund Programme (H2020-MSCA-COFUND-2020-101034228-WOLFRAM2). This work benefitted from Dagstuhl Seminar 25072 ``What You Hear is What You See? Integrating Sonification and Visualization.'', the Audible Universe Workshops \cite{misdariis2023audible} and the ICAD 2023 Session on Astronomical Data Sonification (\href{}{https://icad2023.icad.org/astrodatason/}).
\normalsize

\printbibliography

\end{document}